# NMR microscopy for *in vivo* metabolomics, digitally twinned by computational systems biology, needs a sensitivity boost

Jan G. Korvink*, Vlad Badilita, Lorenzo Bordonali, Mazin Jouda, Dario Mager, Neil MacKinnon

Institute of Microstructure Technology, Karlsruhe Institute of Technology,
Hermann-von-Helmholtz-Platz 1, 76344 Eggenstein-Leopoldshafen, Germany

The metabolism of an organism is regulated at the cellular level, yet is strongly influenced by its environment. The precise metabolomic study of living organisms is currently hampered by measurement sensitivity: most metabolomic measurement techniques involve some compromise, in that averaging is performed over a volume significantly larger than a single cell, or require invasion of the organism, or arrest the state of the organism. NMR is an inherently non-invasive chemometric and imaging method, and hence *in principle* suitable for metabolomic measurements. The digital twin of metabolomics is computational systems biology, so that NMR microscopy is potentially a viable approach with which to join the theoretical and experimental exploration of the metabolomic and behavioural response of organisms. This prospect paper considers the challenge of performing *in vivo* NMR-based metabolomics on the small organism *C. elegans*, points the way towards possible solutions created using MEMS techniques, and highlights currently insurmountable challenges.

1. **Introduction**

"Metabolomics" is the study of the molecules of life, as expressed by the reagents and products of bio-chemical reactions taking place within the cells of a living organism. The open access KEGG pathway database[1,2] describes the known metabolomic pathway maps for a range of organisms, linking and organising contributions from the scientific literature. An organism's metabolome is a time-dependent fingerprint of the state of a cell, and full knowledge of the metabolome would in principle reveal many upstream mechanisms at the cellular level that form part of the omic chain (genomics, transcriptomics, proteomics, etc), and link them with downstream behavioural responses. It is a key premise of systems biology[3] that the metabolome can be computed via a partial differential equation system for the joint metabolomic pathways, and practitioners aim to use the results to predict organism response to disease or environmental influences. To date, no general *non-invasive* method exists with which to measure the *instantaneous metabolome* to sufficient resolution in space, time, species, or rate. In fact, the metabolomic literature is rife with mass spectrometry measurements, which completely destroys the sample by vapourisation, and may require extensive pretreatment such as chromatography. Nuclear Magnetic Resonance (NMR) is a non-invasive alternative to mass spectroscopy to extract detailed information about the metabolic composition of a target sample. On the one hand, the powerful analytical capabilities of NMR derive from the large availability of specialised pulse sequences and methods that allow for extremely granular investigation of a metabolic profile. On the other hand, the technique suffers from inherently limited sensitivity, which render the feasibility of even simple experiments progressively more challenging as the sample volume (or metabolite concentration) is reduced towards the µL or nL range (µM or nM concentrations). Hence a methodology is currently lacking with which to verify the predictions of systems biology, the digital twin of experimental metabolomics and behavioural studies. In this paper, the need for this experimental capability is detailed, and some solutions are presented that hold promise to enable *in vivo* NMR metabolomics at the single organism level with cellular resolution.

*Corresponding author: e-mail: jan.korvink@kit.edu

We decided to focus our attention on the nematode *Caenorhabditis elegans*.[4] The many reasons to do so lean on those which originally justified Sydney Brenner to select the worm as a basis for rational genomics research:[5]

- *Standardisation*. The embryonic development of *C. elegans* follows an identical map, so that cellular predecessors are exactly known. Responses are also largely programmable, so that a worm colony can be reasonably synchronised.
- *Tractability*. With only ~$10^3$ cells, and 302 neurons, localisation is readily related to function, and to behaviour.
- *Transparency*. Phenotypes are not always optically distinguishable, so that molecular phenotyping takes on a particular significance.
- *Physical dimensions and other practicalities*. The worm (see next section) has about the dimensions of practical microfluidics, and represents approximately the smallest organism size where inductive NMR still has a scaling advantage. It has a very short lifecycle, allowing fast experimentation, and produces large amounts of progeny. The worms can be maintained at -80 °C, and revived easily.

2. **NMR signal strength in the context of metabolomics**

The signal-to-noise-ratio of the NMR experiment (see Section 4) is a fundamental quantity that describes the level of signal that can be obtained from the sample, given the specific hardware choices, such as the size and arrangement of the detector, the polarization field, and of course the sample type and concentration. Our purpose is to determine the signal intensity per unit sample volume, in units that are easy to scale and assess. We consider the adult nematode *C. elegans* throughout, which has a length of about ~1100 µm. The worm has a widest diameter of about 80 µm yielding a cross-sectional area of ~5000 µm$^2$, and hence a volume of ~5 pL per µm of body length. The total body volume is therefore ~5 nL. Typical metabolite concentrations lie in the mM to nM range, from which we can determine the available signal intensities. For example, in *C. elegans*, a concentration of 1 nM (µM, mM) corresponds to ~$3.3 \times 10^6$ molecules ($10^9$, $10^{12}$), or ~$10^3$ per cell ($10^6$, $10^9$). For an NMR limit-of-detection of ~$10^{12}$ molecules, this implies requiring ~$10^6$ worms ($10^3$, $10^0$) for a distinguishable signal.

As we will see below, the best current NMR technology requires at least an entire worm to measure an NMR signal (Figure 1), and this will yield data for only the most abundant of metabolites, such as the major participants of central glycolysis and the Krebs cycle. This represents a reasonable starting point for *in vivo* metabolomics, since about 1/3 of the worm's body mass is muscle tissue. As we will also see below, it is possible to enhance the NMR signal, and these measures to improve the detectability (or reduce the lower limit of detection) can subsequently be used for either:

- *An extension of the number of detectable metabolites*. Thus more rare metabolites may become detectable, which in turn would reveal the dynamics of metabolomic cycles that involve these metabolites.
- *A reduction of the compartment size*. In this way, we may move from a detectable volume the size of

a worm, to a slice, and perhaps to a cell or organelle. One would then be able to decipher a metabolomic rate at the level of this smaller compartment.

- *A reduction of the time resolution*. If the detection of a well-resolved NMR signal together with signal recovery takes $\tau$ seconds, then $1/\tau$ is the frequency at which we can determine a new concentration of a specific metabolite, so that, following the Nyquist theorem, we can resolve the dynamics of the metabolite to at most a frequency of $1/(2\tau)$. For $\tau$ we have $\tau > 2 \cdot min\ (T_1, T_2)$.

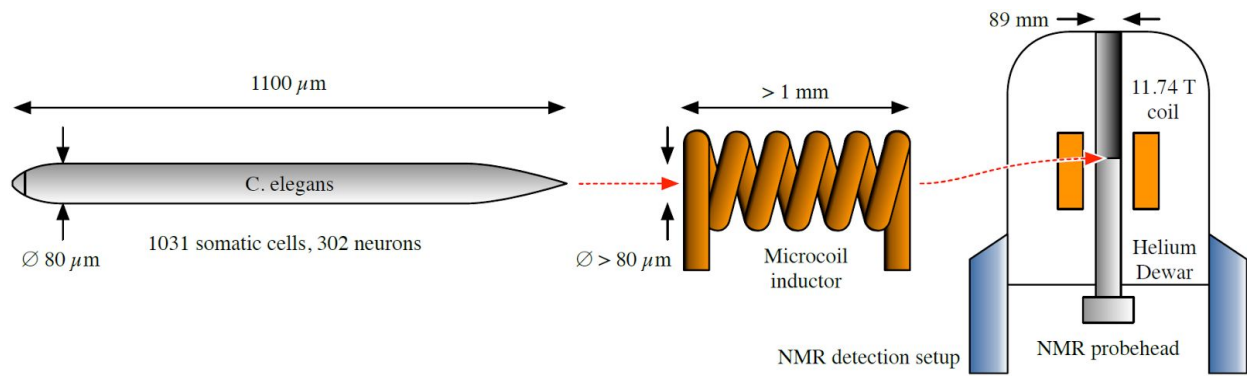

Figure 1. The *C. elegans* worm is placed within the windings of a microcoil resonator, in the strong field of an NMR magnet, for subsequent NMR spectroscopy or imaging.

3. **Types of NMR experiments**

The magnetic resonance frequency $\omega = 2\pi \cdot f$ of an ensemble of nuclei with gyromagnetic ratio $\gamma$ depends on the strength of the magnetic field $B_o$ they are subjected to through the relation $\omega = \gamma \cdot B_o$. The frequency is primarily determined by the strength of the NMR magnet, which is built to have a highly uniform field distribution, so that any remaining differences in frequency only pertain to the sample, and not the hardware. Sample induced changes in magnetic field strength are caused by the local chemically defined magnetic environment in a particular molecule, referred to as chemical shift, a value that can be very precisely determined for a particular chemical species. It is also possible to impose a coordinate dependent magnetic field gradient, which then enables the spatial encoding of the signal. Further signals can be generated by correlating two or more frequencies mediated by inter-nucleus couplings, which can be used to make very precise assignments of molecular structure and dynamics (see Spin Dynamics[6] for details). Thus magnetic resonance spectroscopy (MRS) primarily focuses on frequency-based discrimination of nuclei and structural elucidation, and magnetic resonance imaging (MRI) primarily obtains spatial distributions of spin density, although both approaches are often combined. Solid samples require special treatment, because the lack of molecular tumbling causes resonance line broadening. Fast spinning of a solid sample at the magic angle w.r.t. $B_0$ has a similar effect of reducing the unwanted dipolar-dipolar coupling term, and poses a challenge for spectroscopy on living samples.

4. **The signal-to-noise ratio of microdetectors**

The signal-to-noise ratio of the inductive NMR experiment was first derived by Hoult *et al.*[7] and is compactly repeated here for convenience. The measurable NMR signal is a time dependent electromotive force generated across the terminals of the NMR detection coil, and produced by an excited ensemble of nuclear spins whose net precessing magnetisation is

$$M(t) = N \cdot \gamma^2 \cdot \hbar^2 \cdot I(I+1) B_0 / (3 k_B \cdot T)$$

where N is the number of resonating spins per unit volume, $\gamma$ is the gyromagnetic ratio of the nuclei of interest, $\hbar$ is the reduced Planck's constant, $I$ is the spin quantum number, $B_0$ is the static magnetic field, and $T$ is the sample's temperature. Besides these parameters, the strength of the NMR signal also depends on the volume of the detection coil, and on the coupling strength between the coil and the magnetisation, or in other words, the filling factor of the coil. This dependence is governed by the reciprocity principle,[7] which states that the time-dependent electromotive force generated across the terminals of the detection coil by the magnetization $\hat{M}(x,t)$ is directly proportional to the field per unit current that specific coil is able to produce for that specific value of magnetization. From this we can derive the relation for the signal for a coil with uniform $B_1$:

$$\xi(t) = \omega_0 \cdot B_{1(xy)} \cdot M_0 \cdot V_s \cdot \cos \omega_0 t$$

where $\omega_0$ is the Larmor precessing frequency, $B_{1(xy)}$ is the transverse field per unit current of the detection coil, and $V_s$ is the sample volume. The coil of length $l$, resistivity $\rho$, permeability $\mu$, and wire perimeter $p$, has a finite radio-frequency resistance $R_c = (l/p)\sqrt{\mu \cdot \mu_0 \cdot \omega_0 \cdot \varrho(T_c)/2}$, so that we can immediately postulate a Johnson/Nyquist noise voltage $\eta(t)$, which represents the amplitude of voltage fluctuations that are observable for an unbiased coil within a specific bandwidth $\Delta f$:

$$\eta(t) = \sqrt{4 k_B \cdot T_c \cdot \Delta f \cdot R_c}$$

The signal-to-noise ratio therefore describes the extent to which a useful signal extends above the background noise signal, i.e., $SNR = \xi(t)/\eta(t)$. This formula sets the boundaries for the parameters that one can optimize in order to obtain higher SNR.

5. **Signal detection**

NMR signals of small samples can be detected in numerous ways, including variously arranged microcoil resonators, stripline resonators, SQUIDs, magnetic resonance force microscopy probes, and nitrogen vacancy centres in diamond. Here we consider practical Faraday coil designs that are suitable for the worm geometry. Uniform fields can be achieved by solenoid, Helmholtz, or saddle coil arrangements of the microwires. To achieve more SNR through miniaturization, we note that the $B_1$-field scales inversely with the coil diameter $d$, i.e., $B_1 \propto 1/d$, hence we target uniform fields from coil windings placed close to the sample. The idea of signal improvement through miniaturisation has been around for quite a while. In 1979 Hoult and Lauterbur laid the foundation by theoretically deriving the scaling effects on the SNR.[8] In 1995 Olson *et al.* used these scaling laws to build a hand-wound solenoidal coil with an inner diameter of 370 μm and hence obtain better or faster results on small (nl) samples.[9] This paper set the upper sensitivity limit for many years since there were no feasible technological alternatives available that could beat this tiny hand-wound coil. Over the years our group among others developed a large variety of MEMS derived techniques. These highly versatile and precise technologies were modified to the needs of 3D coils, examples are the automatic wirebonder,[10] or a process called rolled-up-MEMS[11] where a flexible substrate can be rolled-up to yield a circular coil geometry. Examples of these are shown in Figure 2, as previously published. [11-15]

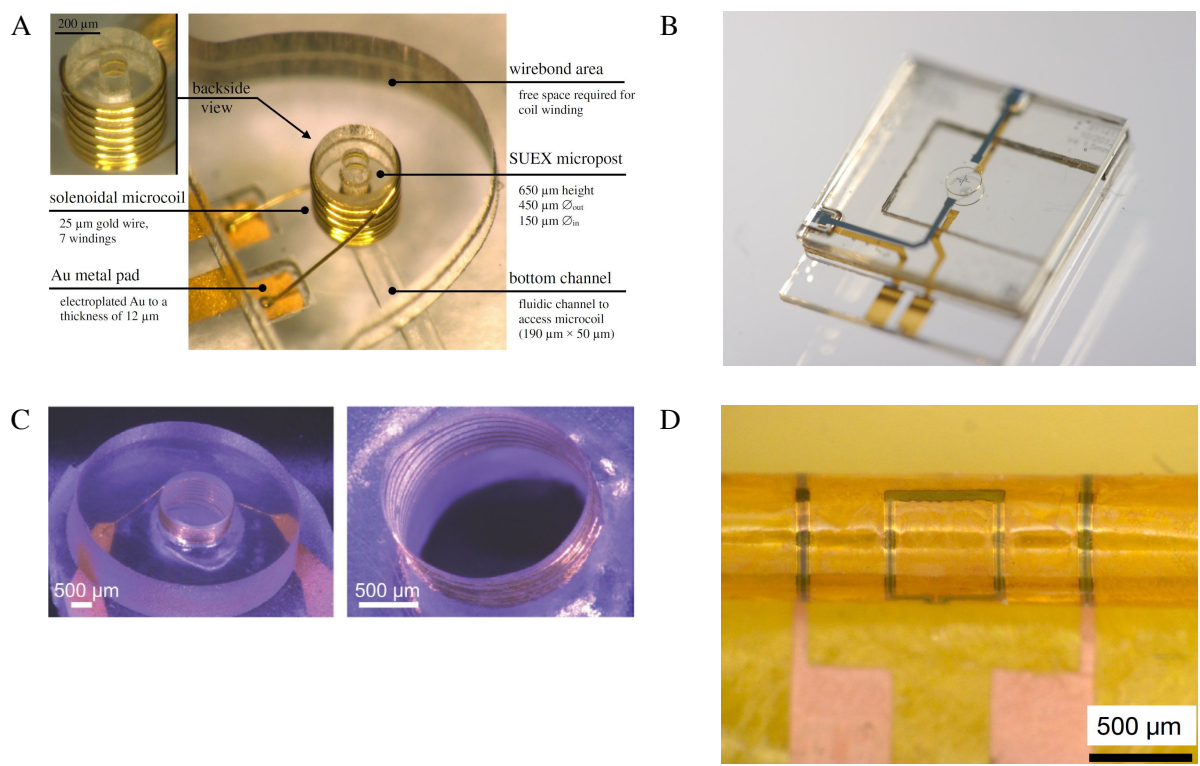

Figure 2. Four microcoil arrangements. A) Solenoid (© IOP Publishing. Reproduced with permission. All rights reserved).[12] B) Helmholtz.[14,15] C) Hollow solenoid.[13] D) Saddle.[11]

Having these technologies allow the fabrication of highly defined microcoils for a huge diameter range down to ∼ 300 μm. With this toolbox at hand the improvement of SNR in microcoils becomes again more of a numerical optimization or design problem than a technological challenge.

6. **Signal enhancement**

The following techniques, which we group here for convenience, can be used to enhance the NMR signal-to-noise ratio, thereby achieving a higher limit of detection:

- *Using stronger magnets*. Normally, the NMR signal is induced by the strong field of a superconducting magnet. By doubling the field strength, say from $B_0 = 11.74$ T to $\alpha_B \cdot B_0 = 23.58$ T, the NMR signal enhancement will be $\alpha_B^{7/4} = 3.36$.
- *By cooling*. At least for small NMR samples, the noise is mainly produced by the detection hardware. Through cooling the electronics to $\alpha_T \cdot T$, the SNR can be improved by a factor $\alpha_T^{-1/2}$. For example, by cooling from room temperature to liquid nitrogen would already enhance the SNR by 1.88. In NMR microscopy it is however challenging to isolate a living sample from closely placed cold hardware.
- *By oversampling*. Multiple NMR signals are added up. For $n$ signals added up, the increase in SNR is proportional to $n^{1/2}$. However, the measurement then takes $n$ times longer, so that the time resolution of the experiment is sacrificed.
- *Through miniaturization.* As we have discussed, reducing a coil's diameter by a factor $\alpha_d$ results in an SNR improvement of $\alpha_d^{-1}$. For example, compared to a 5 mm NMR detector, a microcoil with 0.5 mm diameter brings an SNR enhancement of 10. Radiofrequency phenomena such as the skin and proximity effect make it impractical to miniaturize detector coils below 200 μm diameter and wire thicknesses below 25 μm.

In practise, all of the the above enhancement factors can be combined, but even in combination they bring only mild improvements to the *C. elegans* metabolomics experiment. More dramatic improvements are potentially available through hyperpolarisation, and especially two techniques show promise in the context of the worm:

- Parahydrogen induced polarisation in the form of signal enhancement by reversible exchange (SABRE).[16] In this modality, a source of ground state para-$H_2$ is needed, and is easily achievable at sufficient quantity with a cryogenic setup. At room temperature, hydrogen has an equilibrium spin state distribution of 1/4 para and 3/4 ortho, while at liquid helium temperatures the equilibrium is completely shifted in favour of the para state. The pure para-$H_2$ represents a vast source of exploitable spin polarisation. Using a suitable catalyst, this spin order can be transferred from para-$H_2$ into a desired substrate molecule. Hyperpolarised metabolites can then be tracked by NMR as they metabolised by an organism, revealing their site and rate of consumption.[17] Achievable signal enhancements with SABRE are in the range 10-1000, depending on experimental details. Recent progress in the research on SABRE has also shown the potential for quantitative chemosensing in the micromolar and nanomolar metabolite concentration scales.[18] In the context of miniaturised *in situ* and *in vivo* NMR, the main challenge for implementation of para-$H_2$ signal enhancement consists in providing adequate amounts of para-$H_2$ to sustain continuous hyperpolarisation throughout the timespan of the NMR experiment.
- Dynamic nuclear polarization (DNP).[19] By polarizing the electrons of radicals in a sample, dipole-dipole coupling and spin diffusion can be used to spread the polarisation outwards from the absorption site and across to the sample's nuclei. At least two DNP variants are interesting and conducive to miniaturization:
  - In low field Overhauser DNP, the sample is maintained at room temperature in the liquid state, whilst being subjected to microwave radiation, to achieve polarization enhancements of $^{13}C$ of up to 1300.[20] An interesting variant relies on the nitrogen vacancy centre in diamond, which has a spin-coupled optical transition, and so can be optically pumped. This avoids the use of radicals, but requires NV centres within a few nanometres of the diamond's surface, which is in contact with the sample. Polarization enhancements of up to 2 orders of magnitude can be expected.[21]
  - In high field dissolution DNP, the sample is maintained at a very low temperature as a spin glass, subjected to microwave radiation, and then rapidly thawed to physiological temperatures before being introduced into the organism.[22] Polarization enhancements of up to 50,000 have been reported.[23]

7. **Signal localisation using strong gradients**

The chemical shift in the resonance of a particular nucleus, which is due to the molecular environment,

can be modified to achieve spatial encoding. In essence, the resonance frequency of the nucleus $\omega = \gamma(B[x_0] + x \cdot \partial B/\partial x)$ is modified by a gradient term which is under experimental control. Gradient coils, arranged in three orthogonal directions, can be used to apply a strong magnetic field gradient that either selects a particular resonant slice, or to spatially encode the NMR signal. In this way, magnetic resonance images of a particular atomic nucleus are facilitated, with the image contrast reflecting one of a range of parameters, such as the spin density, or any of the various relaxation times. In metabolomics, slice selection is very useful in order to achieve a smaller compartment size, and to eliminate signals from background materials. Furthermore, gradients can be used in clever ways to achieve signal suppression of unwanted contributors, such as the water background, or other abundant but uninteresting molecules.

8. **Mixture analysis**

When performing NMR on a *C. elegans*, all nuclei within the detector volume will normally contribute to the acquired signal, unless special precautions are taken. A naively recorded 1D NMR spectrum will therefore be an overlay of all the peaks from a very large number of molecules, including all lipids, proteins, DNA, and RNA. Additionally, the organism will most likely be in a nurturing environment, whose components will also contribute to the background signal. To assign the spectrum, that is, to identify the molecules contributing to the signal, the spectral overlap must be addressed. The spectral complexity is reduced simply because resonances emanating from molecules whose abundancies lie below the detection limit will appear as noise. For the remainder of the mixture contributing to the spectrum, special techniques are required to identify the components. In addition to the 1D NMR spectrum, multi-dimensional homonuclear (e.g. $^1$H-$^1$H) or heteronuclear (e.g. $^1$H-$^{13}$C) correlation spectra can be measured in order to reduce spectral overlay as signals are spread into multiple frequency dimensions. Preliminary molecular identification can then be done by comparison to metabolite spectral databases (HMDB,[24] BMRB,[25] MMCD,[26] COLMAR[27]), followed by confirmation by authentic standard spiking. Further development in mixture analysis includes, as an example, extracting spectral assignment from molecular spectral fingerprints derived from time-dependent coherence transfer within the molecule.[28]

9. **Conclusions**

In order to assess future solution strategies, the current limit-of-detection that is preventing the *in vivo* metabolomic monitoring of *C. elegans* requires qualification. For the single most abundant metabolite in the organism, we currently require the signal from at least one nematode. This means that less abundant metabolites or smaller compartments are not yet measurable with NMR. Only through massive signal oversampling at the expense of time resolution, this signal can be enhanced and hence localised into an interesting smaller compartment of the nematode, for example in a slice, or a single cell. We can therefore surmise that we are currently at the horizon of achieving a useful and hence biologically relevant result in terms of SNR and localisation. We have also learned that the inductively-detected small-sample NMR experiment has reached its practical sensitivity limit, with very little hope for dramatic improvement in SNR through further miniaturization or redesign.[29] Any improvement attempt must therefore either consider another more sensitive detection principle altogether, or the introduction of hyperpolarisation methods into biological samples. For example, a factor of $10^3$ in signal enhancement will render mM concentrations single-shot-detectable at the single cell level in *C. elegans*, but µM levels will still require raw oversampling factors of $10^6$, which is only feasible if sparse sampling is possible.

Our main conclusion therefore is that the NMR microscopy of *C. elegans* definitely holds promise for fully resolved *in vivo* metabolomic profiling of an important model organism, but to do so will require at least an additional three orders of magnitude in signal detectability, most likely achieved by using hyperpolarization.

**Acknowledgements**

The work leading to this paper was primarily supported by the European Research Council (ERC) under grant number 290586 (NMCEL). Additional support was provided by the Deutsche Forschungsgemeinschaft (DFG), in the framework of the German Excellence Initiative under grant number EXC 1086 (BrainLinks-BrainTools), and grant number KO1883/23-1 (RUMS). The first author is greatly indebted to his hard-working team and collaborators who supported or contributed the numerous results that are summarized in this publication. Besides the co-authors, these include: Shyam Adhikari, Natalia Bakhtina, Erwin Fuhrer, Andreas Greiner, Oliver Gruschke, Jürgen Hennig, Jens Höfflin, Jan Hövener, Robert Kamberger, Ronald Kampmann, David Kauzlaric, Sebastian Kiss, Mona Klein, Kai Kratt, Robert Meier, Markus Meissner, Ali Moazenzadeh, Nikolaus Nestle, Kirill Poletkin, Herbert Ryan, Pedro Silva, Christoph Trautwein, Marcel Utz, Ulrike Wallrabe, Nan Wang, Peter While, Maxim Zaitzev.

<div align="center">**References**</div>

1. M. Kanehisa, S. Goto: Nucleic Acids Research **28**(2000) 27.
2. M. Kanehisa, S. Goto, Y. Sato, M. Kawashima, M. Furumichi, M. Tababe: Nucleic Acids Research **42**(2013) D199.
3. H. Kitano: Nature **420**(2002) 206.
4. *WormBook:*. The *C. elegans* Research Community, WormBook, http://www.wormbook.org (accessed June 2017).
5. S. Brenner: Genetics **77**(1974) 71.
6. M. H. Levitt: Spin Dynamics: Basics of Nuclear Magnetic Resonance (John Wiley & Sons, 2008).
7. D. I. Hoult, R. E. Richards: J. Magn. Reson. **24**(1976) 71.
8. D. I. Hoult, P. C. Lauterbur: J. Magn. Reson. **34**(1979) 425.
9. D. L. Olson, T. L. Peck, A. G. Webb, R. L. Magin, J. V. Sweedler: Science **270**(1995) 1967.
10. K. Kratt, V. Badilita, J. G. Korvink, U. Wallrabe: J. Micromech. Microeng. **20**(2010) 015021.
11. N. Wang, M. V. Meissner, N. MacKinnon, D. Mager, J. G. Korvink: J. Micromech. Microeng. *Accepted.*
12. R. Ch. Meier, J. Höfflin, V. Badilita, U. Wallrabe, J. G. Korvink: J. Micromech. Microeng. **24**(2014) 045021.
13. R. Kamberger, A. Moazenzadeh, J. G. Korvink, O. Gruschke: J. Micromech. Microeng. **26**(2016) 065002.
14. N. Spengler, A. Moazenzadeh, R. Ch. Meier, V. Badilita, J. G. Korvink, U. Wallrabe: J. Micromech. Microeng. **24**(2014) 034004.
15. N. Spengler, J. Höfflin, A. Moazenzadeh, D. Mager, N. MacKinnon, V. Badilita, U. Wallrabe, J. G. Korvink: PLoS ONE **11**(2016) e0146384.
16. R. W. Adams, J. A. Aguilar, K. D. Atkinson, M. J. Cowley, P. I. P. Elliott, S. B. Duckett, G. G.

R. Green, I. G. Khazal, J. López-Serrano, D. C. Williamson:  Science **323**(2009) 1708.
17. N. M. Zacharias, H. R. Chan, N. Sailasuta, B. D. Ross, P. Bhattacharya: J. Am. Chem. Soc. **134**(2012) 934.
18. N. K. J. Hermkens, N. Eshuis, B. J. A. van Weerdenburg, M. C. Feiters, F. P. J. T. Rutjes, S. S. Wijmenga, M. Tessari: Anal. Chem. **88**(2016) 3406.
19. A. Abragam, M. Goldman: Rep. Prog. Phys. **41**(1978) 395.
20. M. D. Lingwood, S. Han:  J. Magn. Reson. **201**(2009) 137.
21. C. O. Bretschneider, Ü. Akbey, F. Aussenac, G. L. Olsen, A. Feintuch, H. Oschkinat, L. Frydman:  ChemPhysChem **17**(2016) 2691.
22. R. E. Hurd, Y.-F. Yen, A. Chen, J. H. Ardenkjaer-Larsen: J. Magn. Reson. Imag. **36**(2012) 1314.
23. L. F. Pinto, I. Marin-Montesinos, V. Lloveras, J. L. Munos-Gómez, M. Pons, J. Veciana, J. Vidal-Gancedo:  Chem. Commun. **53**(2017) 3757.
24. D. S. Wishart, T. Jewison, A. C. Guo, M. Wilson, C. Knox, Y. Liu, Y. Djoumbou, R. Mandal, F. Aziat, E. Dong, S. Bouatra, I. Sinelnikov, D. Arndt, J. Xia, P. Liu, F. Yallou, T. Bjorndahl, R. Perez-Pineiro, R. Eisner, F. Allen, V. Neveu, R. Greiner, A. Scalbert:  Nucl. Acids Res. **41**(2013) D801.
25. E. L. Ulrich, H. Akutsu, J. F. Doreleijers, Y. Harano, Y. E. Ioannidis, J. Lin, M. Livny, S. Mading, D. Maziuk, Z. Miller, E. Nakatani, C. F. Schulte, D. E. Tolmie, R. K. Wenger, H. Yao, J. L. Markley:  Nucl. Acids Res. **36**(2007) D402.
26. Q. Cui, I. A. Lewis, A. D. Hegeman, M. E. Anderson, J. Li, C. F. Schulte, W. M. Westler, H. R. Eghbalnia, M. R. Sussman, J. L. Markley:  Nat. Biotechnol. **26**(2008) 162.
27. F. Zhang, S. L. Robinette, L. Bruschweiler-Li, R. Brüschweiler:  Magn. Reson. Chem. **47**(2009) S118.
28. N. MacKinnon, P. T. While, J. G. Korvink:  J. Magn. Reson.**217**(2016 )147.
29. V. Badilita, R. Ch. Meier, N. Spengler, U. Wallrabe. M. Utz, J. G. Korvink:  Soft Matter **8**(2012) 10583.